\documentstyle[fullpage,epsf,twocolumn]{article}
\def\<{\langle}
\def\>{\rangle}
\twocolumn
\title{Edge-driven transition in surface structure of nanoscale silicon}
\author{Sohrab Ismail-Beigi and Tom\'{a}s Arias\\
Department of Physics\\
Massachusetts Institute of Technology\\
Cambridge, MA 02139}

\date{\ }

\begin{document}

\maketitle

\begin{abstract}
We present an {\em ab initio} exploration of the phenomena which will become
important for freestanding structures of silicon as they are realized
on the nanoscale.  We find that not only surface but also edge effects
are important considerations in structures of dimensions $\sim$3 nm.
Specifically, for long nanoscale silicon bars, we find two competing
low-energy reconstructions with a transition from one to the other as
the cross section of the bar decreases.  We predict that this
size-dependent phase transition has a signature in the electronic
structure of the bar but little effect on elastic properties.
\\
\break PACS {\bf 68.35.Bs  68.35.Rh}
\end{abstract}
\\

As our understanding of bulk and surface properties of materials
matures, the physics of nanoscale structures opens new fundamental
questions.  What are the ground state structures of nanoscale
collections of matter and to what extend can they be predicted by
simply scaling down bulk and micron-level behavior or scaling up the
behavior of small clusters?  What new considerations must be taken
into account?  Do nanoscale structures exhibit fundamentally different
electronic or mechanical properties due to the large fraction of atoms
at surfaces {\em and edges}, i.e., at the intersection of two
surfaces?  What are the the effects of nanoscale structure on the
reconstruction of surfaces?

Clearly, for sufficiently small structures, edges become important.
One key issue is the identification of the scale at which this happens
and in particular whether edges come into play for anything larger
than a small cluster of atoms.  Also, one must determine the phenomena
by which this importance manifests itself.  In this letter, we use
{\em ab initio} calculations to show that edge effects indeed become
important in silicon on length-scales on the order of a few
nanometers, only a factor or two or three times smaller than what can
be achieved by recent technology\cite{GaagScherer}.  We find that the
presence of edges has a profound effect on the reconstruction on the
surface of a structure and thereby its electronic structure.
Specifically, we predict that for long bars along the [001] direction
the edges drive a surface reconstruction transition from the familiar
``2x1'' family to the ``c(2x2)'' family\cite{Ihm} at a
cross section of 3\ nm\ $\times$\ 3\ nm.

\begin{figure}
\epsfxsize=3.0in
\epsfbox{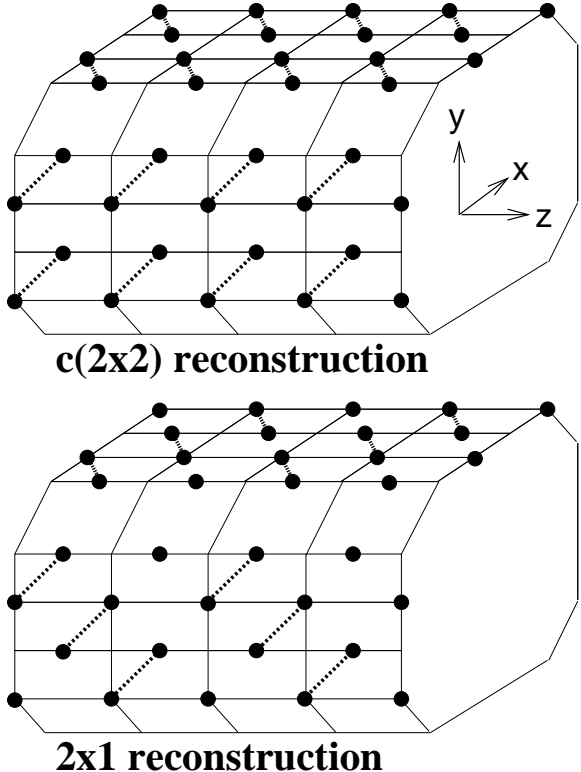}
\caption{A schematic view of the bar and the bonds formed on the
surface of the bar in the c(2x2) reconstruction (top) and in the
2x1 reconstruction (bottom).}
\label{fig:bars}
\end{figure}

Long ``bars'' (as illustrated in Figure \ref{fig:bars}) provide the
ideal laboratory for studying the nature of edges and their
interaction with surfaces.  An isolated edge implies an infinite
system, whereas a bar consists of a series of edges bounding a finite
area and thus may be studied within the supercell framework.  In addition,
such structures are studied experimentally.  Using lithographic
techniques, long bars of silicon can be created in the form of
suspended bridges between bulk silicon
supports\cite{SchererRoukes}. The heat flow and vibrational properties
of such structures should be unique, reflecting quantum confinement
and quantization of bulk phonons\cite{SchererRoukes}. Furthermore,
since the initial report of bright visible luminescence from ``porous
silicon''\cite{Canham}, there have been many efforts to explain this
phenomenon based on quantum confinement in silicon wires or
bars\cite{LDAbars,Yorikawa,Hybersten,ZungerWang}.

In this work, we take on the question of determining the ground-state
structure of nanometer sized bars of silicon as a central issue.
Calculations to date, where this has not been the central issue, have
all been done with hydrogen-passivated silicon surfaces, placing the
atoms at their ideal bulk coordinates\cite{Yorikawa} or simply
relaxing to the closest energy minimum without exploring alternate
constructions\cite{LDAbars}.  Hydrogen-passivation of silicon surfaces
prevents many different types of reconstructions, a subject of
interest when silicon surfaces are exposed to vacuum and which we
study here.  Furthermore, the question of the ``rounding'' of such
bars by the formation of facets along their edges has generally been
ignored.

\begin{figure}
\epsfxsize=3.2in
\epsfbox{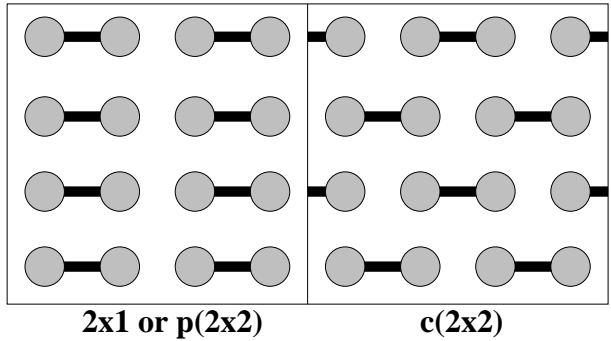}
\caption{These two schematic top views of the Si(100) surface show the
dimerization patterns for both the 2x1 or p(2x2) (left) and c(2x2)
reconstructions (right).}
\label{fig:2x1c2x2schematic}
\end{figure}

{\em Background information---} All of the {\em ab initio} electronic
structure calculations which we report here were carried out within
the total energy plane wave density functional pseudopotential
approach\cite{RMP}, using the Perdew-Zunger\cite{PerdewZunger}
parametrization of the Ceperly-Alder\cite{CeperlyAlder}
exchange-correlation energy and a non-local pseudopotential of the
Kleiman-Bylander form\cite{KB} with $p$ and $d$ non-local corrections.
In all cases, we used a plane wave basis set with a cutoff energy of
12 Ry.  Electronic minimizations were carried out using the
analytically continued functional minimization approach\cite{ACprl}.

To establish baseline information which we shall need later, we
computed {\em ab initio} energies for various reconstructions of the
Si(100) surface.  These calculations were carried out in a supercell
geometry with slabs of twelve atomic layers containing 48 silicon
atoms and separated by 9\AA\ of vacuum.  We sampled the Brillouin zone
using the four {\em k} points $(0,\pm{1 \over 4},\pm{1 \over 4})$.
Among the 2x1, p(2x2), and c(2x2) reconstructions of the (100) surface
(Figure \ref{fig:2x1c2x2schematic}), we find in good agreement with
previous calculations\cite{othersurfcalcs} that the p(2x2) is lowest
in energy with a binding energy of 1.757~eV/dimer, and that the 2x1
reconstruction is higher in energy than the p(2x2) by 0.114~eV/dimer.
Furthermore, we find the c(2x2) reconstruction to be higher in energy
than the p(2x2) by 0.154~eV/dimer. To determine the expected size of
the facets along the edges of the bars we require the Si(100) and
Si(110) surface energies.  These are known experimentally to be 1.36
and 1.43 J m$^{-2}$ respectively\cite{Eaglesham}.

\begin{figure}
\epsfxsize = 2.5in
\centerline{\hfil\epsffile{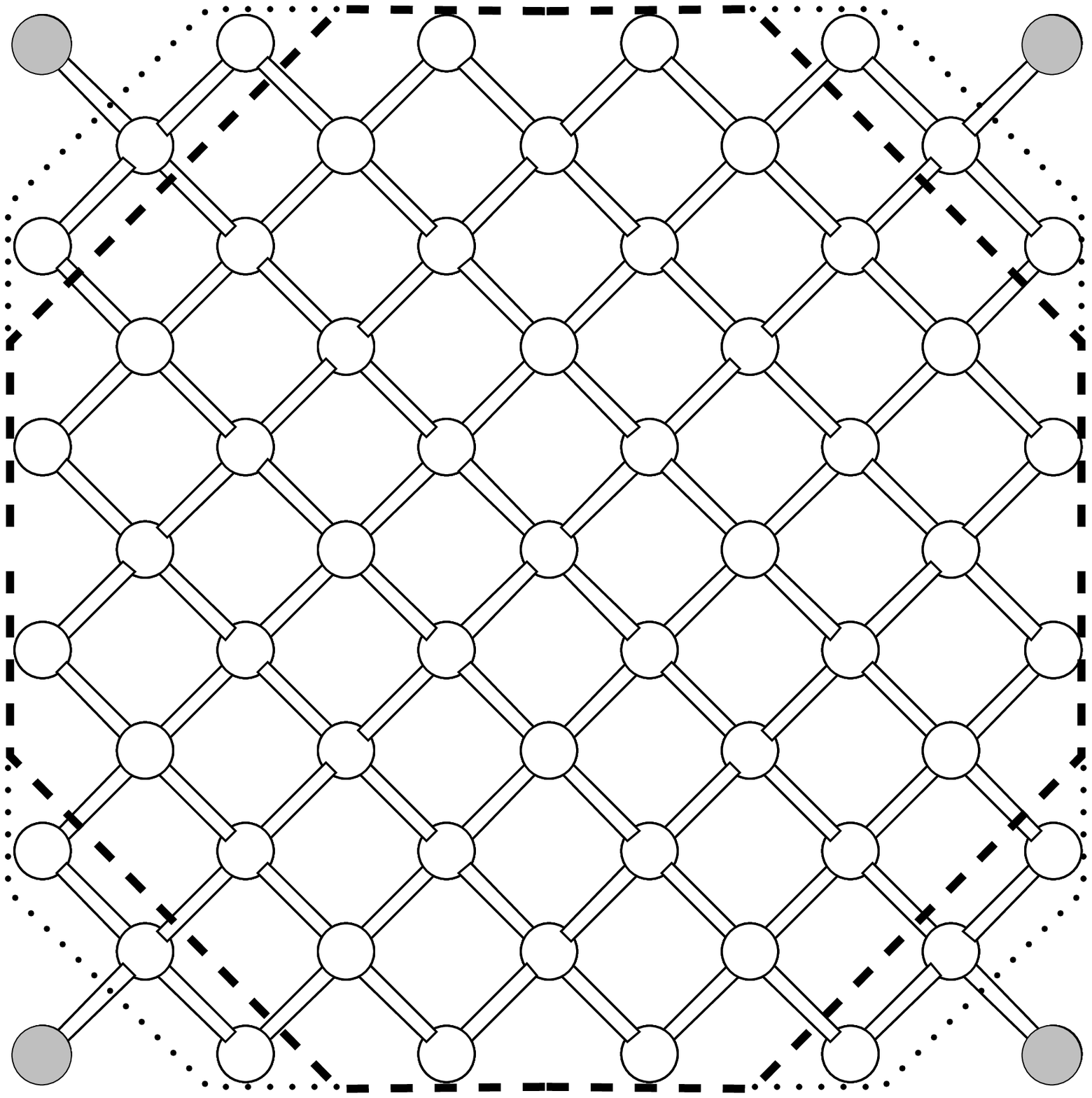}\hfil}
\caption{Cross sectional view of the silicon bars showing the two different
Wulff constructions that result from using either the experimental
(dashed lines) or the tight-binding surface energies (dotted
lines).  The final structure used in this study is formed by removing
the shaded atoms along the edges.}
\label{fig:wulffcuts}
\end{figure}

Our {\em ab initio} study of edges is carried out on bars with a cross
section of $2.5\times2.5$ cubic unit-cells in the (001) plane.  We
apply infinite periodic boundary conditions in the [001] direction with
a periodicity of two cubic unit cells.  The ball-and-stick diagram in
Figure \ref{fig:wulffcuts} depicts the projection of this structure on
to the (001) plane.  {\em Ab initio} calculations on the bars were
performed using the same pseudopotential and energy cutoff as used
above for the surfaces.  For the bars, we sampled the Brillouin zone
at the two {\em k} points $(0,0,\pm{1 \over 2})$ and provided for a
minimum of 6\AA\ of vacuum between periodic images of the bars.

{\em Prediction of a finite-size transition ---} Before determining
the reconstructions along the edges and surfaces, we first must
determine the overall cross-sectional geometry of the bar.  To
establish this, we performed the Wulff construction using the
experimental surface energies given above.  Figure \ref{fig:wulffcuts}
shows the resulting shape when \{110\} facets connect \{100\} surfaces
of silicon, scaled to the lateral size of our bars.  The atomic-scale
structure most consistent with the Wulff construction in shape and
aspect ratio appears in Figure \ref{fig:wulffcuts}, where we have
removed the four columns of shaded atoms along the edges.  The
projection in the (001) plane of the final structure is an octagon.
With the aforementioned periodicity along [001], the final structure
contains 114 atoms.  Without relaxation, all atoms on the surfaces of
the bar are two-fold coordinated.

Having determined the overall geometry, we may now turn to the more
subtle issue of relaxation and possible reconstructions along the
surfaces and edges.  We have found two competing structures: one which
best satisfies the system in terms of the total number of bonds, and
the other which best satisfies the system in terms of the
configuration of the exposed surfaces.  We find that the system cannot
satisfy both conditions simultaneously.

Starting from the unrelaxed configuration, there is one unique surface
atom with which each edge atom may bond in order to become three-fold
coordinated.  This bonding does not change the periodicity along the
edge, which is the [001] vector of the crystal lattice.  This
periodicity, however, is incompatible with the periodicity of the
p(2x2) low energy state of the \{100\} facets.  To maintain maximal
bonding, the arrangement of atoms on the \{100\} facets must then
revert to the higher energy c(2x2) reconstruction.  We denote this
configuration of the bar as the ``c(2x2) reconstruction'' (See Figure
\ref{fig:bars}).

While the c(2x2) reconstruction maximizes the number of bonds in the
system, the increased energy represented by the c(2x2) arrangement on
the \{100\} facets will eventually outweigh the benefit of maximal
bonding along the one dimensional edges for a sufficiently large system.  When the \{100\} facets
assume the p(2x2) configuration, the periodicity along the [001]
direction is doubled.  The edge atoms are no longer equivalent, and
every other atom now cannot form a new bond and remains only two-fold
coordinated.  Because of bonding restrictions into the interior of the
structure, the dimer rows along alternate faces assumes a pattern
which leads us to denote this configuration as the ``2x1
reconstruction.'' (See Figure \ref{fig:bars}.)

The ground state of the bar is thus determined by the balance between
bonding along the edges and the surface energy of the facets.  We find
that for our bar the effects from the edges overcome the natural
tendency of the surfaces.  The c(2x2) configuration is lower in energy
than the 2x1 configuration by 1.94 eV per [001] vector along the
length of the bar.  There is therefore a size-dependent transition in
this system as we increase the lateral dimension of the bar and place
relatively more atoms on the \{100\} facets.  Our calculations place
the crossover point at approximately 5 dimer pairs on each (100)
surface per unit cell along [001], corresponding to a bar with cross
sectional dimensions of approximately 3.0\ nm\ $\times$\ 3.0\ nm, far
larger than the scale of an atomic cluster.  We therefore predict an
important edge-driven transition at a scale only two or three smaller
than what has been achieved experimentally to date\cite{GaagScherer}.

More generally, we can see that the compatibility of the competing
surface reconstructions with the translational symmetry of the edges
plays a pivotal role in determining the ground state of nanoscale
structures.  In our specific case, it is the principal physical
mechanism giving rise to the size-dependent transition for our bars.

\begin{figure}
\epsfxsize=3.2in
\epsfbox{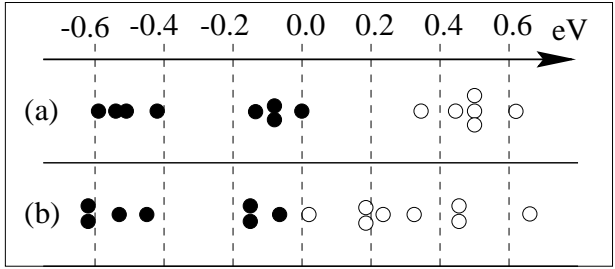}
\caption{Eigen-energies of the electronic states for the two
reconstructions of the bars in the vicinity of their Fermi levels at
the $k$ points used in the calculations: (a) is the 2x1
configuration, (b) is the c(2x2) configuration.  Filled states are
denoted by filled circles and empty states by empty circles.  The zero
of energy is arbitrary.}
\label{fig:barstates}
\end{figure}

{\em Implications of the transition---} Comparing the electronic
structures of the two reconstructions (see Figure
\ref{fig:barstates}), we find that the 2x1 configuration has a gap
of 0.35 eV across the Fermi level.  Furthermore, the topmost filled
states consist of four nearly degenerate states which are localized on
the four edges of the bars, and this cluster is separated from states
below it in energy by a gap of 0.30 eV.  This nearly symmetric
placement with sizeable gaps as well as the spatial localization of
the edge states leads us to conclude that the 2x1 bar is
insulating.

On the other hand, the electronic structure of the c(2x2) configuration
is more subtle.  The states in the vicinity of the Fermi level are
localized on the surfaces of the bars, and the gap across the Fermi
level is only 0.09 eV.  We believe the c(2x2) configuration is a
small-gap semiconductor or even perhaps metallic.  Thus the nature of
the low energy electronic states and excitations differ between the
two bars and should manifest themselves in physical measurements such
as electrical conductivity or optical spectra.

Next, considering possible differences in mechanical properties, we
carried out molecular dynamics simulations using a tight-binding model
to compute transverse acoustic phonon frequencies for the bars.  We
believe that the tight-binding model will give us a qualitative view
of the differences which may exist between the two bars.  If any large
differences are found, we should examine this issue in more detail
using the more demanding {\em ab initio} techniques.

We used the semi-empirical tight-binding model of Sawada\cite{Sawada}
with the modification proposed by Kohyama\cite{Kohyama}. This model
provides a good qualitative description of bulk, dimer, and surface
energetics of silicon.  In using the Sawada model, we only keep
Hamiltonian matrix elements and repulsive terms between atoms closer
than $r_{nn}=6$\AA.

In order to calculate the band-structure energy, we used the fully
parallelizable $O(N)$ technique of Goedecker and
Colombo\cite{GoedeckerColombo}.  By checking the convergence of the
total energy to its ground-state value (determined by exact
diagonalization), we found it necessary to use the following set of
parameters to ensure convergence of energies to within an accuracy of
$10^{-4}$ eV/atom: in the nomenclature of \cite{GoedeckerColombo}, we
have $k_BT = 0.125$ eV, $n_{pl} = 300$, and $r_{loc} = 15.0$ \AA.

Table \ref{table:TB} summarizes the results of the Sawada model for
the various key physical values in this study.  Although the Sawada
model correctly predicts asymmetric dimers as the ground-state of the
Si(100) surface, it is not sensitive to the delicate rocking of dimers
that differentiates the 2x1 and p(2x2) reconstructions and hence finds
the 2x1 reconstruction to be lower in energy than the p(2x2).
However, in our study, the relevant energy difference is that between
the c(2x2) and the lowest-energy reconstruction of the surface, and
this quantity is reproduced rather well.  The Sawada model also does
well in predicting surface energies: Figure \ref{fig:wulffcuts} shows
the result for the Wulff construction when we use the tight-binding
surface energies, and the resulting geometry is very similar to the
experimentally derived one.  Thus we believe that the Sawada model
provides a good semi-quantitative description for the physics of our
bars.

Using the Sawada model, we computed transverse acoustic phonon
frequencies for the $\vec{k} = (\pi a/4)\hat{z}$ mode (along $\Delta$)
for both reconstructions of our bars, the longest allowed wavelength
along the length of the bar consistent with the periodic boundary
conditions.  We ran $(N,V,E)$ molecular dynamics simulations using the
Verlet algorithm with a time-step of 2.4 fs for 900 time steps.
Phonon frequencies were identified as peaks in the frequency-domain
power-spectrum of the velocity autocorrelation function as estimated
by auto-regressive fits.  We found frequencies of $1.98\pm 0.02$ THz
and $1.93\pm 0.02$ THz for the c(2x2) and 2x1 configurations
respectively.  Thus we see that edge effects do not have a significant
effect on the long wavelength vibrations of the bars.

{\em Conclusions---} We have performed an {\em ab initio} study of the
energetics of long bars of silicon in vacuum.  We found useful the
atomic-scale version of the Wulff construction as a first step in
determining nanoscale structure.  Next, we found that one cannot
ignore the interplay between the edges and surfaces in silicon
structures with dimensions of a few nanometers, where the
compatibility of the surface reconstructions with the symmetry of the
edges plays an important role.  In particular, the ground-state of our
bars changes from one surface reconstruction to another, signalling a
cross section-dependent phase transition, with significant influence on the
electronic structure of the system.  Finally, we find that even on the
scale of a few nanometers, the surface-edge interplay has little
effect on mechanical properties.

\begin{table*}
\begin{tabular}{|l|c|c|}
\hline
& Expt/Ab initio & TB \\
\hline
\multicolumn{3}{|c|}{Bulk properties} \\
\hline
Binding energy (eV/atom) & -4.63$^a$ & -4.79 \\
\hline
Bulk modulus (Mbar) & 0.975$^a$ & 0.906 \\
\hline 
\multicolumn{3}{|c|}{Phonon frequencies} \\
\hline
$\Gamma$ (THz) & 15.5$^b$  & 18.1  \\
\hline
$\vec{k} = (\pi /4a)\hat{z}$ LA (THz) & 3.9$^b$  & 3.8  \\
\hline
$\vec{k} = (\pi /4a)\hat{z}$ TA (THz) & 2.4$^b$ & 3.0 \\
\hline
\multicolumn{3}{|c|}{Si(100) reconstructions} \\
\hline
lowest energy (eV/dimer) & p(2x2): 1.757 & 2x1: 2.05 \\
\hline
c(2x2)  (eV/dimer) & 1.603 & 1.93 \\
\hline
\multicolumn{3}{|c|}{Surface energies} \\
\hline
Si(100) (J m$^{-2}$) & 1.36$^c$ & 1.44 \\
\hline
Si(110) (J m$^{-2}$) & 1.43$^c$ & 1.77 \\
\hline
\multicolumn{3}{|c|}{Differences between zizag and c(2x2) bars} \\
\hline
Energy difference (eV/$a$) & 1.94 & 1.57 \\
\hline
Cross-over (width in nm) & $3.0$ & $3.0$ \\
\hline
\end{tabular}
\caption{ Comparison of the Sawada model (TB) with experimental and
{\em ab initio} results: (a) is reference \protect \cite{Kittel}, (b)
is reference \protect \cite{Dolling}, and (c) is reference \protect
\cite{Eaglesham}.  The energy difference between the 2x1 and c(2x2)
reconstructions of the silicon bars are given in eV per unit cell
along [001], and the cross-over refers to the approximate width of a
bar when the two reconstructions have the same energy (see text).}
\label{table:TB}
\end{table*}

\begin{center}{\bf Acknowledgments}\end{center}
We thank Prof. Alan Edelman, Matteo Frigo, Andrew Pochinsky, and Prof.
John Negele for valuable discussions regarding optimization and
parallelization of the computer code.  We also thank Nicolaj Moll for
providing us with Ref \cite{Eaglesham}.  The calculations were carried
out on the Xolas prototype SMP cluster.  This work was supported
primarily by the MRSEC Program of the National Science Foundation
under award number DMR 94-00334 and also by the Alfred P. Sloan
Foundation (BR-3456).

\end{document}